\documentclass[letter]{aa}
\usepackage{txfonts,epsfig}

\begin{document}
\title{The remnant of SN\,1987A revealed at (sub-)mm wavelengths\thanks{Based
on observations collected at the European Southern Observatory, Chile (ESO
N$^{\rm o}$ 78.F-9034).}}
%\subtitle{is something missing?}
\author{Ma\v{s}a Laki\'cevi\'c\inst{1,2}
\and    Jacco Th.\ van Loon\inst{2}
\and    Ferdinando Patat\inst{1}
\and    Lister Staveley-Smith\inst{3}
\and    Giovanna Zanardo\inst{3}
}
%\offprints{M.\ Laki\'cevi\'c}
\institute{European Organization for Astronomical Research in the Southern
           Hemisphere (ESO), Karl-Schwarzschild-Str.\ 2, D-85748, Garching b.\
           M\"unchen, Germany, \email{mlakicev@eso.org}
\and       Astrophysics Group, Lennard-Jones Laboratories, Keele University,
           Staffordshire ST5 5BG, UK
\and       International Centre for Radio Astronomy Research, M468, University
           of Western Australia, Crawley, WA 6009, Australia}
\date{Accepted: 28 April 2011}
\abstract
% context heading (optional)
{Supernova 1987A (SN\,1987A) exploded in the Large Magellanic Cloud (LMC). Its
proximity and rapid evolution makes it a unique case study of the early phases
in the development of a supernova remnant. One particular aspect of interest
is the possible formation of dust in SN\,1987A, as SNe could contribute
significantly to the dust seen at high redshifts.}
% aims heading (mandatory)
{We explore the properties of SN\,1987A and its circumburst medium as seen at
mm and sub-mm wavelengths, bridging the gap between extant radio and infrared
(IR) observations of respectively the synchrotron and dust emission.}
% methods heading (mandatory)
{SN\,1987A was observed with the Australia Telescope Compact Array (ATCA) at
3.2 mm in July 2005, and with the Atacama Pathfinder EXperiment (APEX) at 0.87
mm in May 2007. We present the images and brightness measurements of SN\,1987A
at these wavelengths for the first time.}
% results heading (mandatory)
{SN\,1987A is detected as an unresolved point source of $11.2\pm2.0$ mJy at
3.2 mm ($5^{\prime\prime}$ beam) and $21\pm4$ mJy at 0.87 mm
($18^{\prime\prime}$ beam). These flux densities are in perfect agreement with
extrapolations of the powerlaw radio spectrum and modified-blackbody dust
emission, respectively. This places limits on the presence of free--free
emission, which is similar to the expected free--free emission from the
ionized ejecta from SN\,1987A. Adjacent, fainter emission is observed at 0.87
mm extending $\sim0.5^\prime$ towards the south--west. This could be the
impact of the supernova progenitor's wind when it was still a red supergiant
upon a dense medium.}
% conclusions heading (optional)
{We have established a continuous spectral energy distribution for the
emission from SN\,1987A and its immediate surroundings, linking the IR and
radio data. This places limits on the contribution from ionized plasma. Our
sub-mm image reveals complexity in the distribution of cold dust surrounding
SN\,1987A, but leaves room for freshly synthesized dust in the SN ejecta.}
\keywords{circumstellar matter
-- supernovae: individual: SN\,1987A
-- ISM: supernova remnants 
-- Magellanic Clouds
-- Radio continuum: ISM
-- Submillimeter: ISM}
\authorrunning{Laki\'cevi\'c et al.}
\titlerunning{SN\,1987A at (sub-)mm wavelengths}
\maketitle
%=========================================================================== 1
\section{\label{sec:intro}Introduction}

Supernovae (SNe) and their remnants (SNRs) are important sources of matter and
momentum to a galactic interstellar medium (ISM) and the main creators of
chemical abundances throughout the Universe. The SN explosions of massive
stars are also a potentially important source of dust, with estimates ranging
from 0.2--4 M$_\odot$ of dust (Dunne 2003) to much lower measured amounts
$<0.01$ M$_\odot$ (Sandstrom et al.\ 2009). These kinds of measurements are
complicated, because SNRs are frequently witnessed to hit surrounding dust
clouds (e.g., N\,49 in the LMC -- van Loon et al.\ 2010; Otsuka et al.\ 2010).

The remnant of SN\,1987A in the Large Magellanic Cloud (LMC) is one of the
most studied SNRs thanks to its proximity ($\approx50$ kpc) and early stage of
(rapid) evolution. Its progenitor is believed to have been a red supergiant
(RSG) in a binary system, prior to becoming the blue supergiant that was
witnessed to vanish (Barkat \& Wheeler 1988, and references therein). Its
circumburst environment has been studied in great detail, revealing a dusty
equatorial ring $\sim2^{\prime\prime}$ in diameter (Bouchet et al.\ 2006) with
two fainter isomorphological rings -- but offset from the former -- on either
side of SN\,1987A. The light echo from SN\,1987A has been used to probe the
surrounding ISM (Xu, Crotts \& Kunkel 1995). The blast wave from SN\,1987A
likely reached dense circumburst regions around day 6000 and the resulting
increased radiation is observed at all frequencies (Dwek et al.\ 2010; Zanardo
et al.\ 2010).

While the emission from the immediate vicinity of SN\,1987A at radio
frequencies is due exclusively to synchrotron radiation arising from
magnetized plasma (see Zanardo et al.\ 2010 for an overview), the mid-infrared
(IR) emission is attributed to a few $10^{-6}$ M$_\odot$ of warm ($\sim170$ K)
RSG dust in the equatorial ring (Bouchet et al.\ 2006). As much as $\sim1$
M$_\odot$ of cold ($\sim20$ K) dust has been detected in {\it Herschel}
images, and interpreted as having been freshly-synthesized in the ejecta of
SN\,1987A (Matsuura et al.\ 2011).

The as yet unexplored sub-mm and mm portion of the electro-magnetic spectrum
offers an additional avenue to probe the coldest dust and to possibly reveal
additional sources of radiation. We present here the observations of SN\,1987A
at 0.87 mm and 3.2 mm, constituting the first detections of SN\,1987A at
wavelengths $0.35<\lambda<8$ mm.

%=========================================================================== 2
\section{Measurements}

We retrieved unpublished data obtained with the Atacama Pathfinder EXperiment
(APEX) from the European Southern Observatory (ESO) archive, and re-analysed
data obtained with the Australia Telescope Compact Array (ATCA) that have not
previously appeared in the refereed literature. The position of SN\,1987A is
(RA $5^{\rm h}35^{m}28.0^{\rm s}$, Dec $-69^\circ16^\prime11^{\prime\prime}$).

%------------------------------------------------------------------------- 2.1
\subsection{APEX observations at 0.87 mm}

%
% FIGURE 1
%
\begin{figure}
\centerline{\psfig{figure=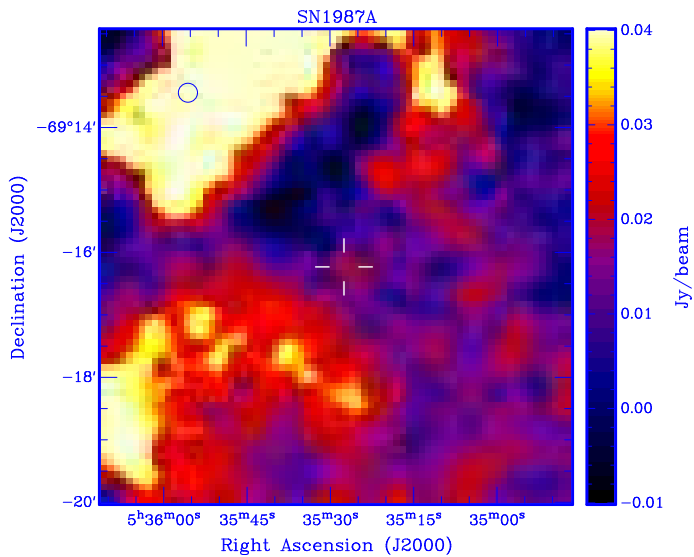,width=90mm}}
\caption[]{APEX image of SN\,1987A at 345 GHz/0.87 mm, obtained in May 2007.
The position of SN\,1987A is indicated with a cross; the beam is drawn in the
upper left corner.}
\label{Apexgauss}
\end{figure}

SN\,1987A was observed with the APEX with the Large Apex BOlometer CAmera
(LABOCA) at a wavelength of 0.87 mm (345 GHz, with a bandwidth of 60 GHz) on
30 May 2007. The total on-source integration time was 2 hr. The field that was
imaged in raster-map mode (to fill the spaces between the 295 bolometer array
elements) subtends $15^\prime$, centred on (RA $5^{\rm h}35^{m}16.8^{\rm s}$,
Dec $-69^\circ15^\prime00^{\prime\prime}$), and the beamsize has a full-width
at half-maximum (FWHM) of $18^{\prime\prime}$.

The data were reduced using the Bolometer array Analysis (BoA) software,
following standard sub-mm calibration procedures. The individual raster-map
scans were first corrected for attenuation by the Earth's atmosphere, signal
calibration factors, and response variations across the bolometer array, and
the voltages were then converted to flux densities using the conversion factor
determined from measurements of planets during commissioning of the APEX.
Then, a series of 20 $\kappa$--$\sigma$ clipping iterations were performed to
remove noise spikes that deviate from the mean by more than
$\kappa\times\sigma$ (with $\kappa\geq4$), taking known bad channels into
account. The individual maps were then combined into a single image. The image
was smoothed with a Gaussian kernel of $\sigma=1$ pixel, i.e.\ an FWHM of
$14^{\prime\prime}$, using the Munich Interactive Data Analysis Software
(MIDAS), and the resolution of this image is thus $23^{\prime\prime}$.

The resulting image is shown in Fig.\ 1. SN\,1987A is clearly detected, though
it sits at the extremity of a fainter, elongated structure extending
$\sim0.5^\prime$ towards the south--west. Part of a molecular cloud complex
fills the North--East corner, while a rim of clouds in the south--east may be
associated with the large Honeycomb SNR\,B0536$-$69.3 (Chu et al.\ 1995). We
used the {\sc center/moment} task to measure the flux density of SN\,1987A
from a Gaussian fit to be $21\pm4$ mJy. Attempts to measure the flux density
using different software suggest an accuracy of $\sim10$\%, but the highly
structured background leads us to adopt a more conservative value of
$\sim20$\%. The MIDAS task that we used also returned a value of
$16^{\prime\prime}$ for the FWHM, which is in fair agreement with the angular
resolution, lending support to the detection of a point source at the position
of SN\,1987A.

%------------------------------------------------------------------------- 2.2
\subsection{ATCA observations at 3 mm}

SN\,1987A was observed with the ATCA on 11--12 July 2005, with the 3-mm
receiver in two frequency bands, one between 93.45--93.56 GHz (peak reception
at 93.49 GHz) and another between 95.5--95.6 GHz (peak reception at 95.57
GHz). The observations were carried out with the most compact antenna
configuration, H75, with an on-source observing time of $\sim8$ hr.

The bands were reduced simultaneously using standard procedures within the
{\sc miriad} software, and ultimately merged. Firstly, the signal was
corrected for elevation-dependent attenuation using up-to-date antenna
reconfiguration files with the task {\sc atfix}. The phase stability and
antenna tracking were investigated and found to be mostly within 0.2--0.5 mm
and (typically) 0.3--$0.6^{\prime\prime}$ to (occasionally)
$\sim3^{\prime\prime}$. Data obtained during tracking errors of
$>3^{\prime\prime}$ were flagged as bad, with the task {\sc uvflag}. Secondly,
the task {\sc mfcal} was used to calibrate the bandpasses and phase offsets
against observations of the primary reference sources 1921$-$293 (at the
start) and 1253$-$055 (at the end), and the phase variations against regular
observations of the secondary reference sources 0637$-$752 and 0454$-$810.
Absolute flux calibration was achieved using the task {\sc mfboot} against an
observation of the planet Uranus. The flux calibration appeared reliable at a
10--20\% level. Thirdly, the signal as function of baseline in the $uv$ plane
(visibilities) was inverted using a standard fast-Fourier transform with the
task {\sc invert}, where the visibilities were weighted according to the noise
variance (i.e.\ system temperature) but otherwise uniformly across a
1-arcminute field chosen to effectively suppress the sidelobes. A
multi-frequency synthesis approach was taken, producing an image at 94.5 GHz
(3.2 mm) with an r.m.s.\ noise level of 2.0 mJy beam$^{-1}$. This map was
cleaned with the tasks {\sc clean} and {\sc restor}, employing a H\"ogbom
scheme (H\"ogbom 1974). The resulting beam has an FWHM of
$5.0^{\prime\prime}\times4.7^{\prime\prime}$ in RA and Dec.

%
% FIGURE 2
%
\begin{figure*}
\centerline{\hbox{
\psfig{figure=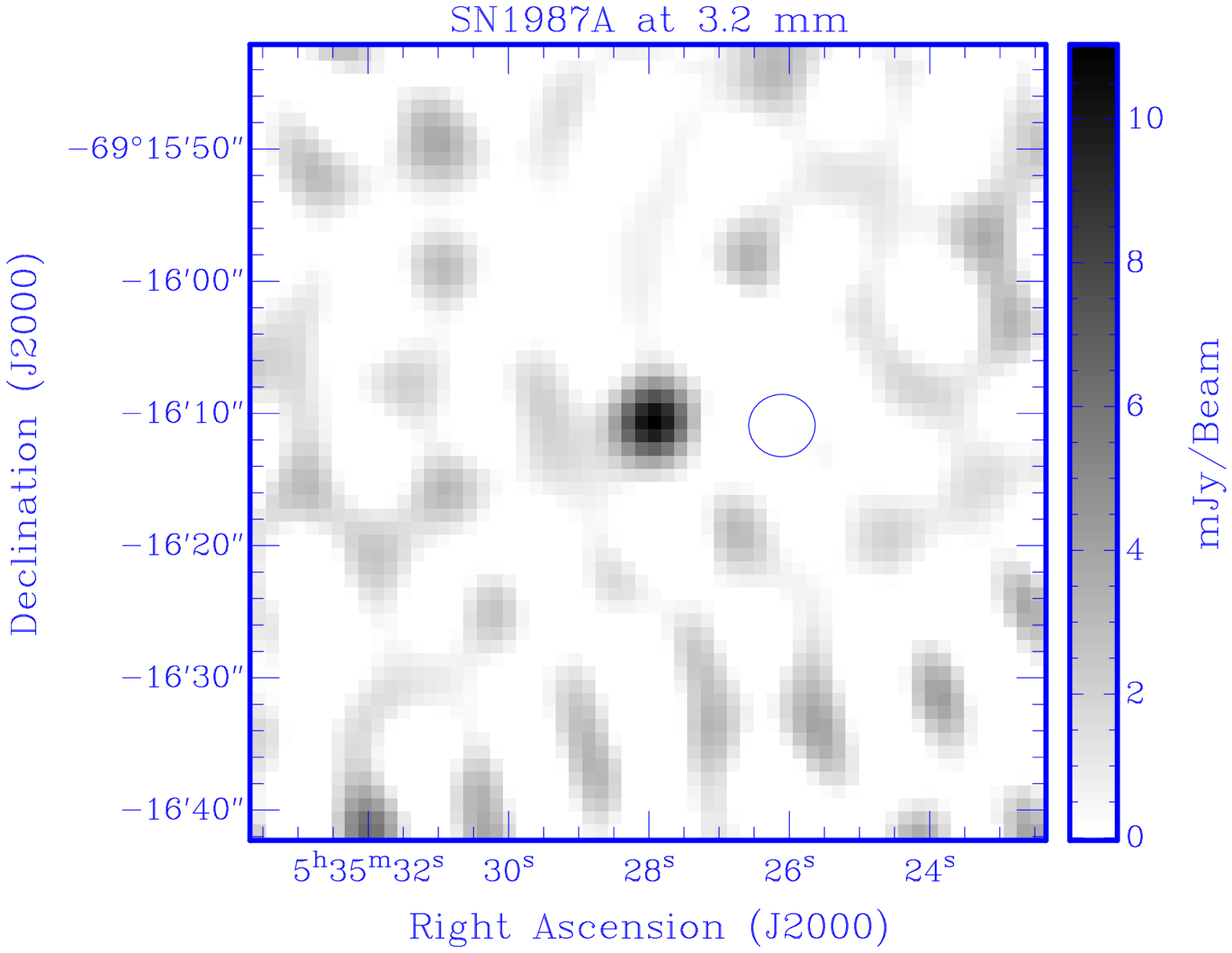,width=84mm}\hspace{5mm}
\psfig{figure=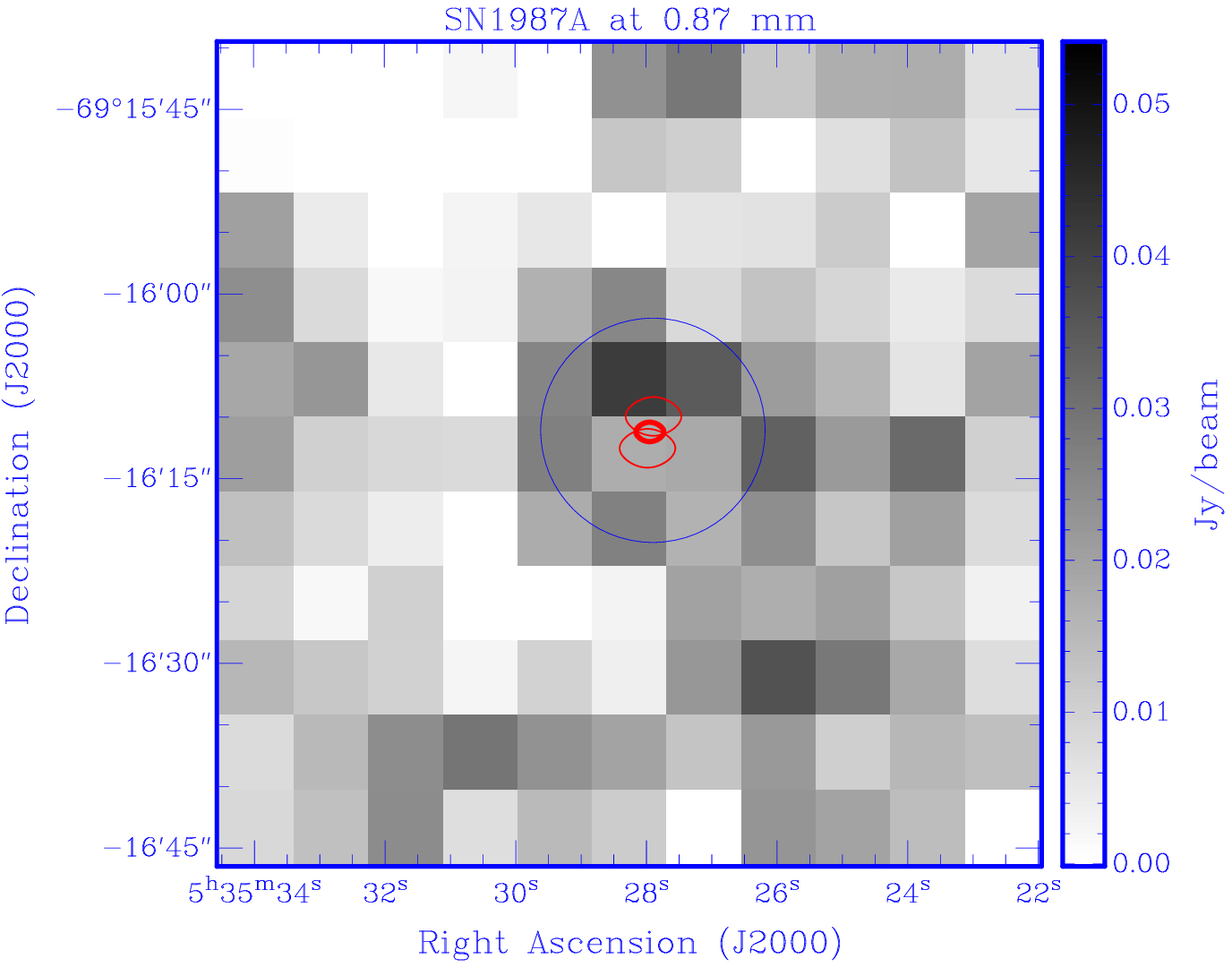,width=88mm}
}}
\caption[]{{\it Left:} ATCA image of SN\,1987A at 94.5 GHz/3.2 mm, obtained in
July 2005. The synthesized beam is drawn just next to it. {\it Right:}
Unsmoothed APEX image of SN\,1987A at 345 GHz/0.87 mm, shown on the same scale
as the 3.2 mm image. The beam is drawn centred on the exact position of
SN\,1987A; the inset is a schematic of the ring structures seen in {\it HST}
images.}
\label{SN1987A3mm}
\end{figure*}

The calibrated image is shown in Fig.\ 2 (left), where SN\,1987A is clearly
detected as an unresolved source at a level of $11.2\pm2.0$ mJy. The other
intensity peaks in the image are much dimmer and unlikely to be real
astrophysical sources.

%=========================================================================== 3
\section{Discussion}

%------------------------------------------------------------------------- 3.1
\subsection{Bridging the IR and radio spectral energy distributions}

%
% modified figure 3 to include a panel below with the residuals from the fits:
%
%
% FIGURE 3
%
\begin{figure}
\centerline{\psfig{figure=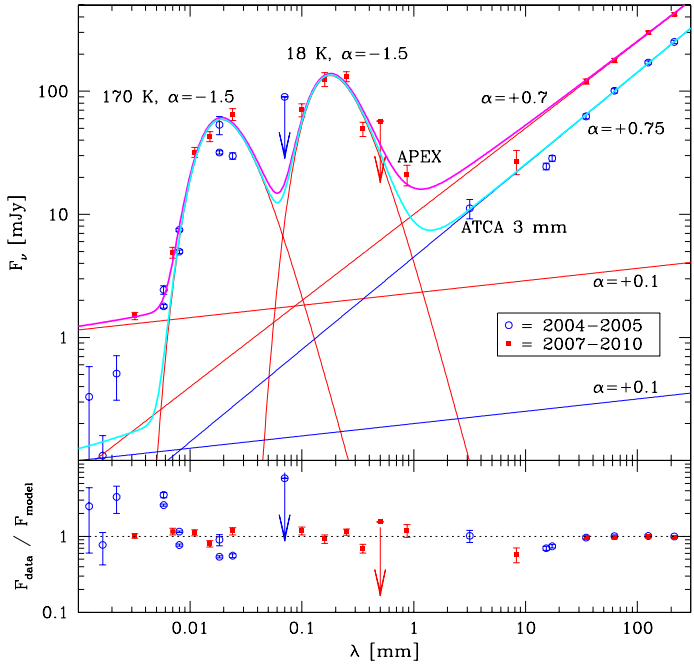,width=90mm}}
\caption[]{Spectral energy distribution from the IR (Bouchet et al.\ 2006;
Seok et al.\ 2008; Dwek et al.\ 2010; Matsuura et al.\ 2011) to the radio
(Manchester et al.\ 2005; Potter et al.\ 2009; Zanardo et al.\ 2010),
including our first measurements at sub-mm (APEX) and mm (ATCA) wavelengths.
Overplotted are modified blackbodies to represent two dust components, two
steep radio powerlaws to represent synchrotron radiation, and two shallow
powerlaws to represent free--free radiation. Cyan and magenta curves represent
the sum of all emission for the 2004--5 and 2007--10 period, respectively,
with residuals plotted underneath.}
\label{sn1987a}
\end{figure}

The APEX 0.87 mm and ATCA 3.2 mm detections bridge a gap of a factor $>20$ in
wavelength. They conjoin the far-IR emission detected up to $\lambda=350$
$\mu$m (Matsuura et al.\ 2011) attributed to thermal radiation from dust
grains, on the one hand, and the radio emission detected down to $\lambda=0.8$
cm (Potter et al.\ 2009) attributed to synchrotron radiation from free charge
in a magnetic field, on the other. The full spectral energy distribution (SED)
is plotted in Fig.\ 3, where the IR data comprise groundbased near- and
mid-IR, and {\it Spitzer} observations published by Bouchet et al.\ (2006),
AKARI observations published by Seok et al.\ (2008), a {\it Spitzer} 70-$\mu$m
upper limit published by Dwek et al.\ (2010) and {\it Herschel} observations
published by Matsuura et al.\ (2011)\footnote{The 500-$\mu$m datum is an upper
limit.}. The radio data solely comprise ATCA observations, published by
Manchester et al.\ (2005), Potter et al.\ (2009), and by the most up-to-date
summary at the longest wavelengths by Zanardo et al.\ (2010). In Fig.\ 3 we
group these measurements into either of two sets, as close as possible in time
to the APEX or ATCA 3.2-mm observations, respectively. This is because both
the IR (Dwek et al.\ 2010) and radio (Zanardo et al.\ 2010) brightness has
been observed to be steadily increasing in recent years. The {\it Herschel}
measurements were only taken in 2010, three years after the APEX measurements.

Concentrating on the (sub)-mm region of interest, the far-IR--sub-mm emission
is represented very well by a modified Planck curve of thermal dust emission:
\begin{equation}
F_\nu^{\rm IR}\propto B_\nu(T)\lambda^{\alpha_{\rm IR}},
\end{equation}
for a dust temperature $T=18$ K and coefficient $\alpha_{\rm IR}=-1.5$. The
mid-IR emission can also be matched with thermal dust emission by employing an
identical $\alpha_{\rm IR}$ but different $T=170$ K and a total emitting grain
surface of $10^{-5}$ that of the cold component. While more detailed models
for the mid- and far-IR emission have been presented elsewhere (Bouchet et
al.\ 2006; Dwek et al.\ 2010; Matsuura et al.\ 2011), this parameterization is
consistent with those results and enables a simple decomposition of the
overall SED.

The mm--dm radio emission is very well matched with a powerlaw with a
coefficient typical of synchrotron radiation:
\begin{equation}
F_\nu^{\rm radio}\propto\lambda^{\alpha_{\rm radio}}.
\end{equation}
A value for $\alpha_{\rm radio}\approx0.75$ during 2004--2005 produces an
excellent fit to the 3--21 cm data, as well as our new measurement at 3.2 mm,
a decade higher in frequency. This renders the measurements by Manchester et
al.\ (2005) somewhat ``too'' faint. Likewise, for the period 2007--2010 a
value for $\alpha_{\rm radio}\approx0.7$ fits the 3--21 cm data well, but
again the higher frequency datum published by Potter et al.\ (2009) lies below
an extrapolation of this powerlaw component. Given the difficulty in radio
interferometry of conserving flux at the longest baselines, we are tempted to
believe the 3.2-mm datum is consistent with a pure synchrotron component. In
particular, we do not feel the need to invoke either mm emission from dust
(cf.\ Bot et al.\ 2010) or free--free emission from a hot plasma.

With regard to the latter, we can set meaningful limits on the contribution
from free--free emission:
\begin{equation}
F_\nu^{\rm f-f}\propto\lambda^{\alpha_{\rm ff}},
\end{equation}
where $\alpha_{\rm ff}\approx0.1$, unless it becomes optically thick -- but
this typically happens at wavelengths $\gg$ cm or it reaches the high-energy
limit -- but that is not the case at IR--radio wavelengths. In fact, Fig.\ 3
demonstrates that the free--free contribution is limited by the near-IR data
(that may include contributions from hot dust and/or atomic line emission),
though in recent years it may have become more pronounced and have reached
levels that become significant precisely at mm wavelengths. The excellent
description of the SED by the cold-dust and synchrotron components leaves room
for only a fraction of the measured flux density at 0.87 mm to arise from any
additional component; in the most extreme case, $\sim2$ mJy at 1 mm is caused
by free--free emission. Since the 0.87 mm datum is positioned on the
well-constrained tail of the modified Planck curve, there is little room for
variations in its slope to affect the estimation of the limit on free--free
emission, Matsuura et al.\ (2011) consider temperatures up to 22 K but in
combination with different $\alpha_{\rm IR}$.

To gain a notion of the physical implications of this limit on the free--free
component, we present an example calculation here. Radiation transfer through
a plasma can be written as
\begin{equation}
I_\nu=\int_0^{\tau_\nu}B_\nu(T)\,{\rm e}^{-\tau_\nu}\,{\rm d}\tau_\nu,
\end{equation}
where the optical depth is
\begin{equation}
\tau_\nu=8.24\times10^{-2}T^{-1.35}\nu^{-2.1}E_{\rm c},
\end{equation}
where the continuum emission measure $E_{\rm c}=\int n_+n_{\rm e}\,{\rm d}s$
with the density $n$ measured in cm$^{-3}$ and the pathlength $s$ measured in
pc. The solution to the radiation transfer equation is
\begin{equation}
T_{{\rm b},\nu}=\int_0^{\tau_\nu}T\,{\rm e}^{-\tau_\nu}\,{\rm d}\nu
               =T\left[1-{\rm e}^{\tau_\nu}\right]
\end{equation}
in the isothermal case (constant $T$), where the brightness temperature is
defined as $T_{{\rm b},\nu}=c^2I_\nu/(2\nu^2k)$. At high frequencies,
$\lim_{\tau_\nu\rightarrow
0}T_{{\rm b},\nu}=T\tau_\nu$. Considering 1 M$_\odot$ of ionized ejecta within
a region of dimensions $\sim0.1$ pc, i.e.\ a density $n_+\sim n_{\rm
e}\sim10^4$ cm$^{-3}$, and a temperature $T\sim10^{4-6}$ K, one obtains an
estimated $T_{\rm b,300\,GHz}\sim0.1$ K. With the ejecta subtending
$\sim0.1^{\prime\prime}$ on the sky, this corresponds to $F_{\rm
300\,GHz}\sim1$ mJy. This has a very similar magnitude to the limit we set on
the free--free component. The shock-ionized plasma mass associated with the
equatorial ring is estimated to be only $\sim0.06$ M$_\odot$ at densities of
$10^{3-4}$ cm$^{-3}$ (Mattila et al.\ 2010), and its free--free emission is
thus expected to be two orders of magnitude fainter than in the above
scenario.

%------------------------------------------------------------------------- 3.2
\subsection{The location and origin of cold dust around SN\,1987A}

The APEX images at 870 $\mu$m and {\it Herschel} images at 250 $\mu$m have an
identical resolution, i.e.\ superior to that of the {\it Herschel} images at
350 and 500 $\mu$m (beamsizes of $25^{\prime\prime}$ and $36^{\prime\prime}$,
respectively). While we clearly confirm that most of the detected cold dust
($T\approx18$ K) appears to be concentrated within a region $\ll
23^{\prime\prime}$ (5 pc) across, our APEX image reveals diffuse emission to
the south--west of SN\,1987A, which is not visible on the {\it Herschel}
images, but which coincides with an emission filament seen in {\it HST} images
(cf.\ Fig.\ 1 in Matsuura et al.\ 2011). We dub this the ``S--W plateau''. We
also note that in the unsmoothed APEX image (Fig.\ 2, right), the brightest
pixels are slightly offset to the north--east from the exact position of
SN\,1987A\footnote{Curiously, the orientation of this symmetry is very similar
to that of the ejecta seen in recent {\it HST} images.}. These observations
leave open the possibility that the 18-K emission arises from part of a larger
dust complex, in radiation equilibrium with the ambient interstellar radiation
field.

A shell of 4.5 pc radius surrounding SN\,1987A was postulated by Chevalier \&
Emmering (1989) to explain a light echo signal. They suggest that it marks the
place where the wind of the red supergiant progenitor is decelerated by the
pressure in the surrounding medium. Crotts et al.\ (2000) detected [N\,{\sc
ii}] emission from a region that would mark the overlap between this shell and
the S--W plateau. They also detected emission from a region at the opposite
side of SN\,1987A, and Doppler shifts of the line emission suggested the shell
expands at a rate of 10--15 km s$^{-1}$, commensurate with the wind of a
metal-poor red supergiant (cf.\ Marshall et al.\ 2004). We may have detected
emission from the same structures as seen in [N\,{\sc ii}], with the APEX at
0.87 mm. That no such structures are seen in the more sensitive ATCA image at
3.2 mm suggests that it is more likely to be dust emission ($\alpha<0$) than
free--free emission ($\alpha>0$).

That said, the flux density associated with the point source at the position
of SN\,1987A is consistent with the 18-K dust component conjectured by
Matsuura et al.\ (2011) to be due to freshly-synthesized dust in the ejecta
from SN\,1987A. To show this we employed a modified blackbody of spectral
index $\alpha_{\rm IR}=-1.5$, which is typical of interstellar dust and
therefore does not require ``exotic'' types of grains or ``anomalous''
emission mechanisms.

%=========================================================================== 4
\section{Conclusions}

The first sub-mm (APEX at 0.87 mm) and mm (ATCA at 3.2 mm) observations of
SN\,1987A that we presented here give a first valuable insight into the dust
distribution that surrounds the more compact remnant area. These data bridge
the gap in the SED between the previously detected far-IR emission from dust
grains (at $\lambda\leq0.35$ mm) and radio emission from synchrotron radiation
(at $\lambda\geq0.8$ mm). They do not deviate from extrapolation of the far-IR
and radio SEDs, respectively, and we set a limit on the contribution from
free--free emission from a plasma, which is close to the estimated
contribution from the ionized ejecta from SN\,1987A. Scheduled high-resolution
3.2-mm observations with the ATCA will improve these constraints.

While the 3.2-mm image ($5^{\prime\prime}$ resolution) is point-like the
0.87-mm image ($18^{\prime\prime}$ resolution) also reveals emission from cold
dust towards the southwest, which may be an interaction region for the RSG
progenitor's wind and a dense medium. The (sub)-mm observations presented here
provide a valuable guide for preparing high-resolution observations with the
Atacama Large Millimeter Array (ALMA), which may settle the question as to the
origin of the cold dust that is detected at far-IR (Matsuura et al.\ 2011) and
sub-mm (this work) wavelengths.

%==============================================================================
\begin{acknowledgements}
We are grateful to Dr.\ Thomas Stanke for advice on the reduction of the
archival APEX data. ML acknowledges studentships from ESO and Keele
University. We thank the referee, John Dickel, for the positive report.
\end{acknowledgements}

\end{document}